\documentclass{ws-procs975x65}

\begin{document}

\title{
Uncovering stellar atmospheres \\
with gravitational microlensing telescopes}

\author{M. Dominik}

\address{University of St Andrews, School of Physics \& Astronomy, \\ North Haugh, St Andrews,
KY16 9SS, United Kingdom
\\E-mail: md35@st-andrews.ac.uk}


\maketitle

\abstracts{
A strong differential magnification over the face of the source star
passing the caustic created by a binary lens star allows to measure its
radial intensity profile with an angular resolution of 20--60~nas
from broad-band photometric observations for
$\sim\,$15 stars per year and to study
its chemical composition from time-resolved high-resolution spectroscopy for
$\sim\,$2 stars per year.}

\section{Broad-band photometry and limb darkening}

\begin{figure}[bt]
\centerline{OGLE 1999-Bulge-23}
\centerline{\epsfxsize=6.1cm\epsfclipon\epsfbox{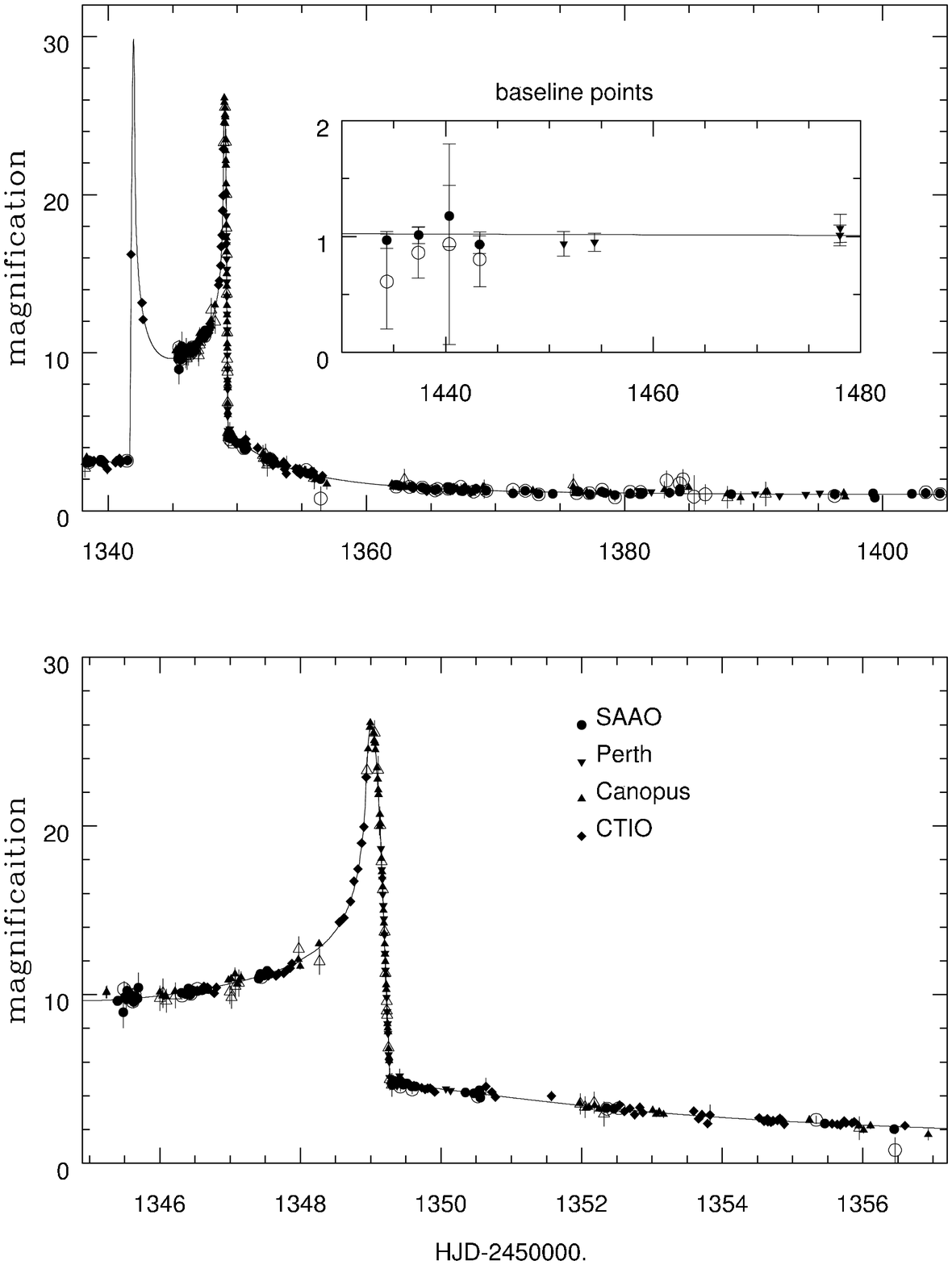}}
\caption{
PLANET $I$-band data from four sites
during the OGLE 1999-Bulge-23 caustic exit.
}
\label{O23}
\end{figure}

During a period of a few hours to a few days, over which a background
source star crosses the caustic due to the gravitational field of a
foreground stellar binary passing close to the line-of-sight,
it encounters a strong differential magnification over its face, 
allowing a measurement of
its radial intensity profile from the dense and precise sampling of the 
microlensing lightcurve which shows a 
characteristic shape.\cite{RB99,SW87}
With a typical proper motion between lens and source star of
10--30~$\mu\mbox{as}~\mbox{d}^{-1}$, a sampling interval of 3~min results in an
effective angular resolution of 20--60~nas for the natural gravitational
telescope.
Using a normalized stellar intensity profile $I_\lambda(\rho)$ corresponding
to a square-root limb-darkening law, 
parametrized as
\begin{displaymath}
I_\lambda(\rho) = 
\left(1-c_\lambda/3-d_\lambda/5\right)^{-1}
\, \left[1+c_\lambda \left(\sqrt{1-\rho^2}-1\right) +d_\lambda
\left(\sqrt[4]{1-\rho^2}-1\right)\right]\,,
\end{displaymath}
where $\rho$ denotes the fractional radius, PLANET has 
published
measurements of limb-darkening coefficients for three K-giants and one
G/K-subgiant in the galactic bulge,\cite{PLANET:M28,PLANET:M41,PLANET:O23,PLANET:EB5}
and together with OGLE, MACHO, MPS, and MOA,
one A-dwarf in the SMC.\cite{joint} 
Table~\ref{thetable} lists the relevant parameters of these events, while
Fig.~\ref{O23} shows the photometric $I$-band data obtained
for the event OGLE 1999-Bulge-23, corresponding to the most precise 
limb-darkening measurement  
of $\sim\,$6\,\%.
The capabilities of current microlensing campaigns allow to determine
up to 15 
limb-darkening coefficients on various types of source stars per year.

\section{Time-resolved spectroscpy and chemical composition}
\begin{figure}[bt]
\centerline{EROS 2000-Bulge-5}
\centerline{\epsfxsize=5.8cm\epsfbox{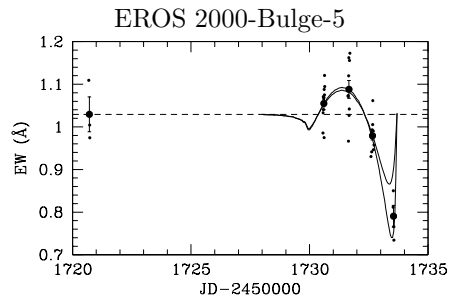}}
\caption{
Equivalence-width variation of H$_\alpha$ during the caustic exit of event
EROS 2000-Bulge-5 from PLANET observations using the FORS1 spectrograph at the VLT.
The big circles with the attached error bars denote nightly averages.
}
\label{EB5}
\end{figure}

According to the variation of the abundance of a chemical element with
stellar radius, the associated spectral lines vary during a caustic
crossing which can be observed by high-resolution spectroscopy providing a
deep probe of the chemical composition of the source star.\cite{GG99}
Using the FORS1 and
UVES
spectrographs at the VLT, PLANET has obtained spectra during caustic
crossings for two microlensing events and observed a significant variation
of the equivalent-width in the H$_\alpha$-line shown
in Fig.~\ref{EB5}.\cite{PLANET:EB5spectra}
With allocated UVES target-of-opportunity time,
PLANET intends to obtain time-resolved spectra during caustic crossings
on 1-2 suitable microlensing events per year.

\begin{sidewaystable}
\tbl{Limb-darkening measurements by PLANET.
\label{thetable}}
{\scriptsize
\begin{tabular}{|lcc|cccccc|}
\hline
\multicolumn{3}{|c|}{}
& \multicolumn{6}{|c|}{}
\\[-1.9ex]
&  &  &  &
 & \multicolumn{2}{c}{MACHO} &  & \\
 & &  & \raisebox{1.5ex}[-1.5ex]{MACHO} & \raisebox{1.5ex}[-1.5ex]{MACHO}
 &\multicolumn{2}{c}{1998-SMC-1}  & \raisebox{1.5ex}[-1.5ex]{OGLE} &
\raisebox{1.5ex}[-1.5ex]{EROS} \\
 & & & \raisebox{1.5ex}[-1.5ex]{1997-Bulge-28} &
\raisebox{1.5ex}[-1.5ex]{1997-Bulge-41} & close binary & wide binary &
\raisebox{1.5ex}[-1.5ex]{1999-Bulge-23} &
\raisebox{1.5ex}[-1.5ex]{2000-Bulge-5} \\[0.35ex] \hline
\multicolumn{3}{|c|}{}
& \multicolumn{6}{|c|}{}
\\[-1.9ex]
\multicolumn{2}{|l}{type of source star} &  & K0\ldots{}K5 III & K III & \multicolumn{2}{c}{A5\ldots{}A7 V} &
G7\ldots{}K2 IV & K2\ldots{}K4 III \\[0.25ex]
\multicolumn{2}{|l}{angular radius $\theta_\star$} &  [$\mu$as] & $8 \pm 2$ & $5.56 \pm 0.54$ &
$0.082 \pm 0.005$ & $0.089 \pm 0.005$ & $1.86 \pm 0.13$ & $6.62 \pm 0.58$ \\[0.25ex]
\multicolumn{2}{|l}{stellar radius $R_\star$} & [$R_\odot$] & $15 \pm 2$ & $10 \pm 2$ & $1.1 \pm 0.1$ &
$1.1 \pm 0.1$ & $3.4 \pm 0.6$ & $12 \pm 3$ \\[0.25ex]
\multicolumn{2}{|l}{source distance  $D_\mathrm{S}$} & [kpc] & $8.5 \pm 1.5$ & $8.5 \pm 1.5$ &
\multicolumn{2}{c}{$60 \pm 2$} & $8.5 \pm 1.5$ & $8.5 \pm 1.5$ \\[0.25ex]
\multicolumn{2}{|l}{proper motion $\mu$} &[$\mathrm{km}\;\mathrm{s}^{-1}\;\mathrm{kpc}^{-1}$] &
$19.4 \pm 2.6$ & $50 \pm 5$ & $1.30 \pm 0.08$ & $1.48 \pm 0.09$ &
$22.8 \pm 1.5$ & $31.1 \pm 2.9$ \\[0.25ex]
\multicolumn{2}{|l}{}  &  [$\mu\mathrm{as}\;\mathrm{d}^{-1}$] & $11.2 \pm 1.5$ &
$29 \pm 3$ & $0.75 \pm 0.05$ & $0.85 \pm 0.05$ & $13.2 \pm 0.9$ &
$18.0 \pm 1.7$ \\[0.25ex]
\multicolumn{2}{|l}{duration entry  $2\,(t_\star^\perp)_1$} & [h] &
$37.4$ & $11.8$ & $6.4$ & $6.6$ & $8.8$ & $26$ \\[0.25ex]
\multicolumn{2}{|l}{duration exit  $2\,(t_\star^\perp)_2$} &  [h] &
$39.2$& $9.2$ & $8.562$ & $8.582$ & $8.194 \pm 0.072$ & $86$ \\[0.25ex]
\multicolumn{2}{|l}{time between c.c.s  $T_{12}$} &  &
$5.08~\mathrm{h}$ & $4.62~\mathrm{h}$ &\multicolumn{2}{c}{$13.2~\mathrm{d}$} & $7.3~\mathrm{d}$ &
$27~\mathrm{d}$ \\[0.25ex]
\multicolumn{2}{|l}{time resolution $\Delta t$} &  [min] & 3 & 10 & \multicolumn{2}{c}{6} & 5 & 15 \\[0.25ex]
\multicolumn{2}{|l}{angular resolution $\Delta \theta$} & [nas] & 23 & 220 & 3 & 3 & 50 & 180 \\[0.25ex]
l.d. coefficients & $c_\mathrm{I}$ &  & $0.40 \pm 0.08$ & $0.52 \pm 0.10$ &
$0.24 \pm 0.05$ (SAAO) & $0.21 \pm 0.05$ (SAAO) & $0.632_{-0.037}^{+0.047}$ & $0.552 \pm 0.090$ \\[0.25ex]
& $c_\mathrm{R}$ & & --- & --- &
$0.24 \pm 0.05$ (EROS) & $0.21 \pm 0.05$ (EROS) & --- & --- \\[0.25ex]
&  &  &--- &--- & $0.06 \pm 0.33$ (CTIO) & $0.06 \pm 0.36$ (CTIO) &
--- & --- \\[0.25ex]
& $c_\mathrm{B}$ &  & --- & --- &
 $0.42 \pm 0.04$ (EROS) & $0.40 \pm 0.04$ (EROS) & --- & --- \\[0.25ex]
&  $c_\mathrm{V}$ & & $0.55 \pm 0.11$ & --- & $0.55 \pm 0.11$ (SAAO) &
$0.50 \pm 0.11$ (SAAO) & $0.786_{-0.078}^{+0.080}$ & --- \\[0.25ex]
& $d_\mathrm{I}$&  & $0.37 \pm 0.07$ &--- &--- &  --- & --- & $0.01 \pm 0.14$ \\[0.25ex]
& $d_\mathrm{V}$&  & $0.44 \pm 0.09$ & --- & --- & --- & --- & --- \\[0.35ex]
\hline
\end{tabular}}
\end{sidewaystable}

\end{document}